\documentclass[preprint]{revtex4}
\usepackage{graphicx}

\begin{document}
\title{Extinction transition on diffusive substrate: a different universality class? }
\author{Galit Evron, David A. Kessler and  Nadav M.  Shnerb}
\affiliation{ Department of Physics, Bar Ilan University, Ramat-Gan
 52900, Israel}

\begin{abstract}
The extinction transition on a one dimensional heterogeneous substrate with diffusive
correlations is studied. Diffusively correlated heterogeneity
is shown to affect the location of the transition point, as the
reactants adapt to the fluctuating environment. At the
transition point the density decays like $t^{-0.159}$, as predicted
by the theory of directed percolation. However, the scaling function describing the behavior away from the transition
shows significant deviations from the DP predictions; it is
suggested, thus, that the off-transition behavior of the system is
governed  by local adaptation to favored regions.
\end{abstract}


 \maketitle

\section{Introduction}

The extinction transition in the stochastic birth-death process is very
important in many branches of science and serves also as a
paradigmatic example of an out of equilibrium phase transition
\cite{1,2,5}. Given a homogeneous substrate and  a single absorbing state,
Grassberger \cite{gras} and Janssen \cite{jan}
conjectured that the microscopic details of the stochastic process
are irrelevant close to the extinction point and the transition is
in the directed percolation (DP) universality class. The basic
rational beyond this conjecture is that a spatially extended  system
splits, close to the transition, into active and inactive zones,
where after each typical period of time there is certain probability
for an active state to die, to survive, or to infect its inactive
neighbors. If these regions are considered as lattice points on a
$d$ dimensional array, the chance of an active site to survive or to
infect its neighbors within a unit time is equivalent to the chance
that a bond exists between lattice point at time $t$ and its
neighbors in a subsequent replica of the system at $t+1$.
Accordingly, the extinction transition happens when the bond density
is exactly at threshold for an infinite cluster, and the transition
belongs to the  directed percolation universality class in  $d+1$
dimensions.

The Grassbrger-Janssen  conjecture has proven to be extremely robust, and
a large number of stochastic models that admit extinction transition
were shown to belong to the DP equivalence class if the substrate is
homogenous \cite{1,2}. It was further shown that  spatio-temporal
substrate noise (i.e.,  birth-death rates that fluctuate in space and time
with only  short range correlations)  is an irrelevant perturbation
close to the transition, so small noise is averaged out and leaves
the DP transition unaffected \cite{1}. Quenched (time independent)
disorder, on the other hand, is a relevant perturbation \cite{jan2}
and seems to  change the nature of the transition. In that case a
Griffiths phase exists between the  active and the inactive regions
\cite{mor}. In the parameter region that corresponds to the Griffiths
phase the survival of an active region depends on the local
properties of the substrate, \emph{not} on activation by neighboring
regions. In particular, for each time scale the survival of active
regions depends on the existence of spatial domains that admit high
carrying capacity \cite{mor,ot}. Although stochastic fluctuations
guarantee extinction for any finite sample, the time scale for that
grows exponentially with the carrying capacity. This implies that
exponentially rare spatial regions with high birth rate support the
population for exponentially large times. An optimization argument
shows that in such a case the survival until $t$ is dominated by
rare spatial fluctuations of linear size $L \sim log(t)$;
accordingly, the density falls algebraically with time. The Griffiths
phase is located between an extinction region, where essentially  no
good islands exist, and the active phase, where good islands infect
each other to yield a never dying process.

What happens, then, if the disorder is neither annealed nor
quenched, but is  diffusively correlated? Two contradictory
arguments may be advanced. In terms of the renormalization group
properties of Reggeon field theory, diffusive disorder is a relevant
operator, so in principle one may expect that the transition will be
in a different equivalence class. This was suggested, in fact, by
Kree, Schaub and Schmittmann~\cite{kss}, who then proceeded  to
develop a perturbative renormalization group based $\epsilon$
expansion around four dimensions, a treatment  that yield
predictions of new (non-DP) critical exponents. One may doubt,
however, the validity of the one-loop approximation used by these
authors below $2d$, as the diffusion constant of the disorder flows
to zero upon successive renormalization group decimation, thus the
perturbative expansion admits only runaway solutions \cite{jan2}.

\begin{figure*}
\includegraphics[width=15cm, angle=270]{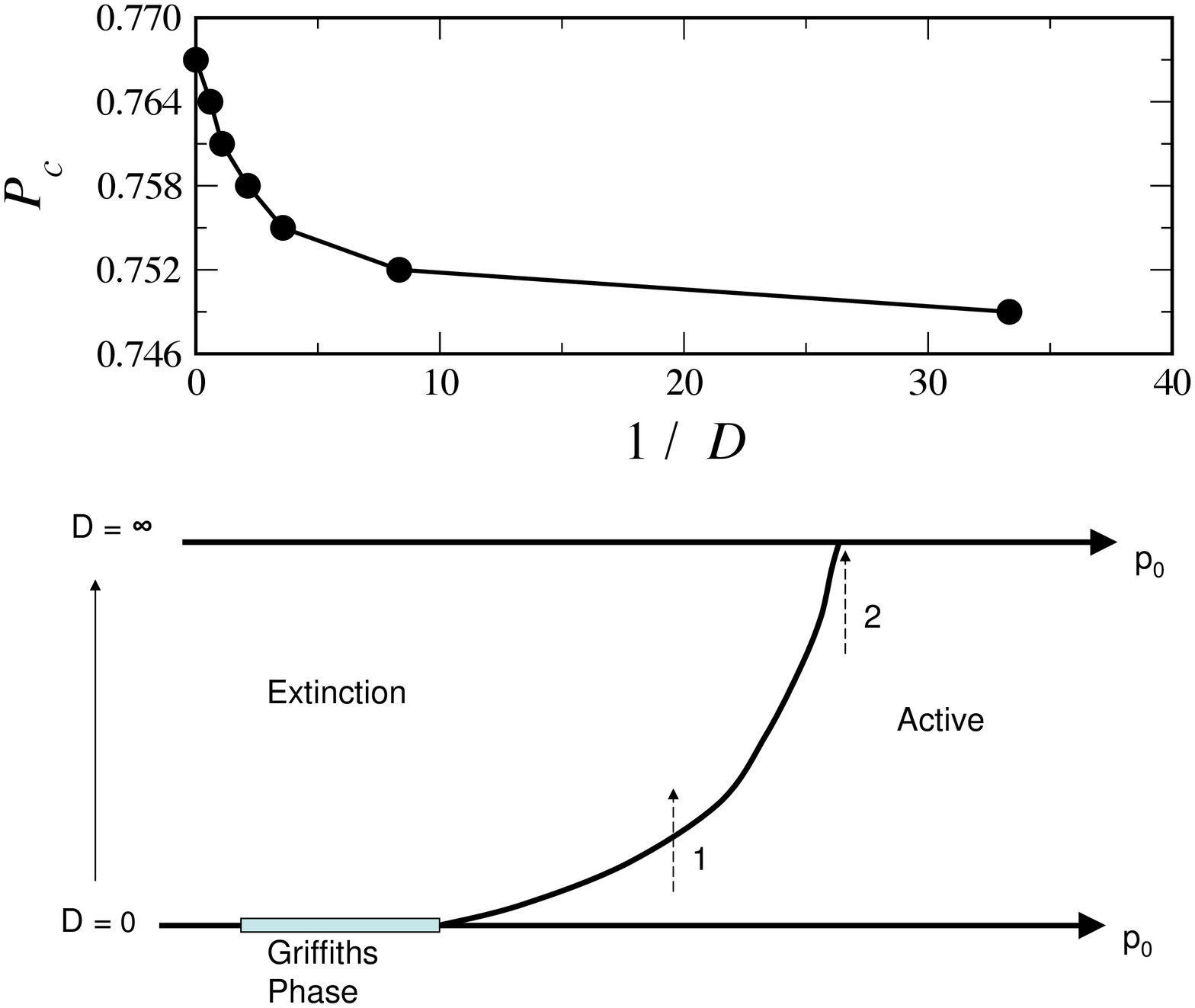}\\
\caption{ A sketch of the phase space in the $D-p_0$ plane (lower
panel), and the actual results measured for the contact process with
a diffusively correlated substrate. As $D$ approaches infinity the
environmental stochasticity becomes uncorrelated in space and time
and the system approaches the homogenous limit of the DP transition.
The critical value of the birth rate, $p_c$, coincides with the
value measures for a contact process on homogenous 1d substrate, as
demonstrated in the upper panel. On the other hand as $D$ become
smaller the effect of local adaptation is significant and the
transition is shifted into the extinction region. As $D \to 0$ the
transition point appears to coincide with the right edge of the
Griffiths phase. In the upper panel the actual values of $P_c$, the
critical values of $P_0$,  are shown. For the same system parameters
the quenched disorder transition (from the Griffiths phase to the
active one) happens at $p_c = 0.746$, a value that corresponds to
the $x$ axis here. Data were obtained from the MC simulation of the
contact process with $L = 10^4$ and $d = 0.1$.}
\end{figure*}

On the other hand, it is clear that the basic intuitive
justification for the Grassberger-Janssen  conjecture is applicable
as long as the favored regions are mobile \cite{yoram}.  Even if the
process is confined to the sites where the diffusive catalysts
exist, these catalysts move randomly in space, thus again any region
may survive, become inactive or infect neighboring regions. One may
suggest, accordingly, that even in the presence of diffusive
catalysts the extinction transition will still be within   the DP
equivalence class.

Here we present  numerical results that suggest a third answer: it
seems that at the transition point the critical exponent that
characterizes the extinction transition is indeed identical with the
DP exponent. However, the subcritical and the supercritical behavior
are not described by the directed percolation scaling function. The
DP transition is characterized by a scaling function of the form
\begin{equation} \label{1}
\rho(t) \sim t^{-\delta} \Phi (\Delta t^{1/\nu_\|})
\end{equation}
Where $\Phi$ is a universal scaling function, $\delta$ is the
universal exponent (in 1d $\delta = 0.159$) $\Delta$ is the distance
from the transition and $\nu_\|$ is the temporal correlation length.
This implies that all the curves coincide when plotting $\rho
t^{\delta}$ vs. $\Delta t^{1/\nu_\|}$. In the presence of diffusive
disorder, while the density decreases like $t^{-\delta}$ where
$\delta$ takes the same numerical value predicted by the DP theory,
the scaling function does not exist, as will be shown below.

The model used here is a contact process  (CP) on a fluctuating
substrate. It take place on a 1d lattice with $L$ sites (with
periodic boundary conditions), where any lattice site is either
occupied by an agent or empty.  In an elementary Monte-Carlo
reaction step an agent is chosen at random and then attempts to
multiply with probability $p$ or "die" (be eliminated from the
sample) with probability $1-p$. If the agent does attempt to
multiply, one of the two neighboring sites in chosen at random and
becomes occupied if it is currently empty; if the site is already
occupied, nothing happens.

To take into account the fluctuating environment, $p$ is taken to be
a space-time dependent fluctuating quantity. At $t=0$, sites are
chosen with equal probability to be either "good" ($p_i = p_0 + d$)
or "bad" ($p_i = p_0 - d$) where $i$ is a site index and $d$ is a
constant.   Subsequently, each elementary step is chosen to be
either a reaction step, as described above, or a catalyst diffusion
step. In a diffusion step, a randomly picked site switches its $p$
value with one of its nearest neighbors. The probability of a given
step to be a reaction step is taken to be $r = \rho/(\rho+D)$, where
$D$ is the catalyst diffusion rate, so that the probability to be a
diffusion step is $1-r$. After each elementary MC step the  time
counter has been advanced by $1/  L (\rho +D ) $.

\begin{figure}
\includegraphics[width=8cm]{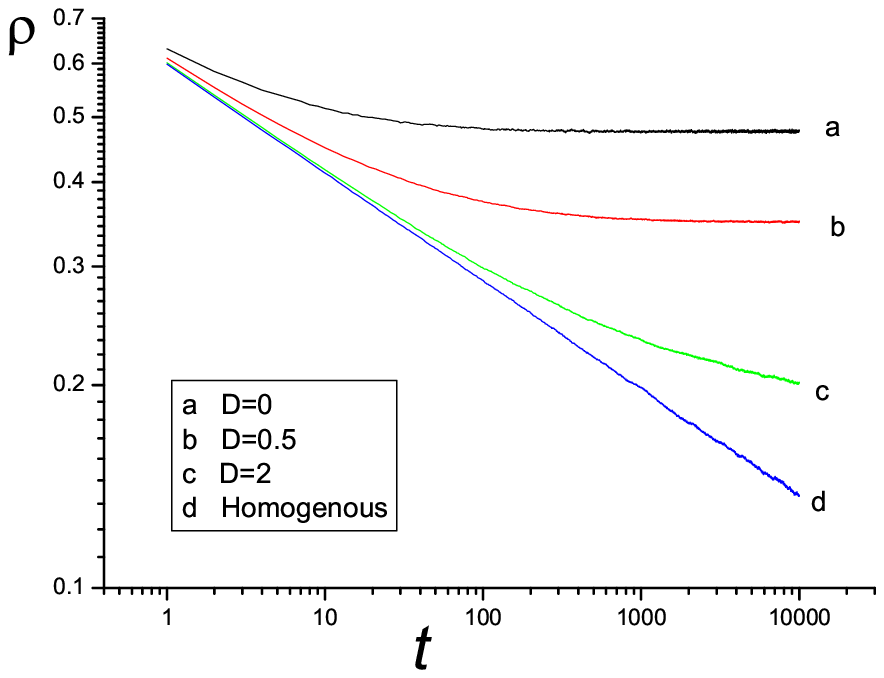}\\
\includegraphics[width=8cm]{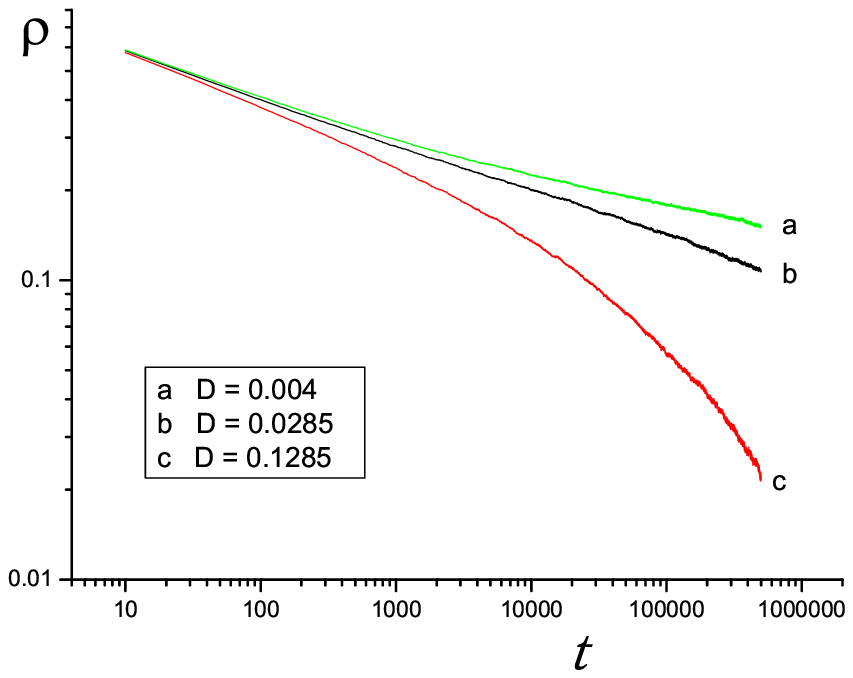}\\
\caption {Density vs. time (logarithmic scale)  at the transition
and in its vicinity. In the upper panel the dashed line (2) in
figure 1 is followed and diffusion is increased with fixed $p_0 =
0.767$ that corresponds to the transition point of the pure contact
process. The system is always in its active phase, as suggested by
the adaptation argument. In the lower panel the separation line is
crossed along arrow (1) of figure 1 ($p_c= 0.749$), and the system
crosses from the active to the inactive phase. In both cases, and
all other cases checked by us (values correspond to the data points
in the upper panel of Fig. 1), the slope at the transition is
$0.159$, in agreement with the DP theory predictions. Datasets were
obtained with lattice size $L=10^4$ and $d=0.1$}
\end{figure}

The phase diagram is shown in Figure 1. For $D \to \infty$, the
space-time disorder becomes uncorrelated and the system belongs to
the DP universality class. At this parameter region the transition
occurs at $p_c = 0.767$, the critical value for the contact process
on a homogenous substrate \cite{5,11}. As $D$ becomes smaller, the
catalysts spend more time in certain spatial regions.  Agents in
these regions produce more offspring in adjacent sites, and when the
catalyst jumps to a neighboring site its probability to be occupied
is larger than average. This implies that reactants actually "adapt"
to the instantaneous configuration of the catalysts, an effect that
yields very strong proliferation in unbounded growth models
\cite{us}. Here the growth is bounded and the effect of adaptation
is finite, still the transition point shifts leftward as exemplified
in Fig. 1. The location of the extinction transition for the slowest
diffusion we were able to measure is very close to the location of
the transition from the Griffiths phase at $D=0$; it seems plausible
that the transition line converges to the right end of the Griffiths
phase as $D$ approaches zero.

Figure 2 shows the approach to the transition from the active and
from the inactive phase. For any crossing of the transition line
checked, either by increasing the diffusion or by changing $p_0$,
the density at the transition decays like a power law with the DP
critical exponent $\delta = 0.159$.  However, the off-transition
behavior does not obey the DP predictions. According to Eq.
(\ref{1}) a graph of  $\rho(t) t^{\delta}$ vs.  $\Delta
t^{1/\nu_\|}$ should be universal close to the transition, as
discussed comprehensively by L\"{u}beck \cite{2}. Figure 3 shows
that this, in fact, the case both on a homogenous substrate and for
infinite diffusion (i.e., when the  locations of the catalysts are
uncorrelated in space and time). On the other hand for finite
diffusion the long-time behavior fails to fit the universal curve;
this implies that a one-parameter  universal scaling function is not
enough to describe the system.

\begin{figure}
\includegraphics[width=9cm]{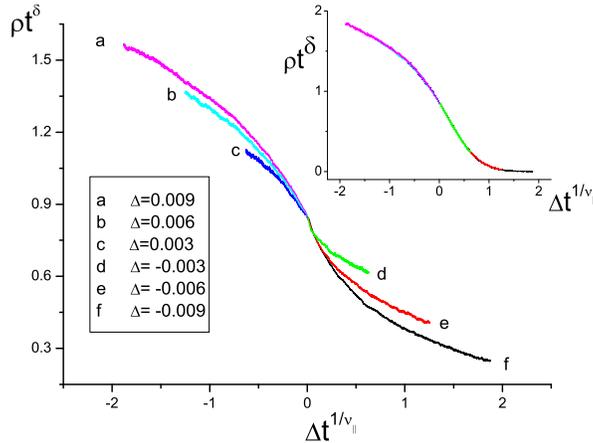}\\
\caption{The universal scaling function for uncorrelated
spatio-temporal  noise (inset) and for finite diffusion ($D =
0.225$) of the catalysts. Different colors correspond to different
distance $\Delta$ from the transition, as explained in the legend.}
\end{figure}

Two types of noise appear in the problem at hand: the demographic
stochasticity associated with the discretness of individual
reactants and the environmental stochasticity associated with the
diffusive wandering of the underlying catalysts. For an unbounded
growth problem (when the carrying capacity of any spatial point is
infinite) this system  is equivalent to KPZ growth with correlated
disorder, but the KPZ perturbative scheme fails for diffusively
correlated disorder  \cite{med} since local adaptation initiates
localized colonies. Indeed, for the unbounded problem extinction
happens almost surely  (i.e., for any finite spatial domain
reactants goes extinct at the long time limit), yet the average
population grows faster than exponentially for $d \le 2$
\cite{kesten}. This effect disappears in  the case of limited
growth, as the adapted colonies can no longer grow exponentially
forever. Still, it seems that the favored regions dominate the
system's behavior above and below the transition. At the transition
point  survival is based on the ability of catalyst rich zones to
"infect" each other, and the microscopic details  average out within
the diverging correlation length. Once $\xi_\bot$ becomes finite
different zones are effectively independent and the spatial
heterogeneity dictates the local decay; in such a case local
adaptation leads to longer survival times and the system behavior
resembles the Griffiths phase, thus leading to deviations from the
DP  universal  scaling function.

\acknowledgements{ The work of N.S.  was supported by the CO3 STREP
of the Complexity Pathfinder of NEST (EC FP6).}

\end{document}